\pdfoutput=1
\documentclass{elsart}
\usepackage{natbib}
\usepackage{graphicx}
\usepackage{amssymb}

\newcommand{\ergpersec}{erg\,s$^{-1}$}

\newcommand{\ergperseccm}{erg\,s$^{-1}$\,cm$^{-2}$}

\newcommand{\ctspersec}{cts\,s$^{-1}$}
\newcommand{\cm}{cm$^{-2}$}

\newcommand{\sgra}{Sgr\,A$^\star$} 
\newcommand{\igr}{IGR\,J17456--2901}
\newcommand{\ax}{AX\,J1745.6--2901}

\newcommand{\grs}{GRS\,1741.9--2853}

\newcommand{\hesssource}{HESS\,J1745--290}
\newcommand{\pwin}{G\,359.95--0.04}

\newcommand{\xmm}{\it XMM-Newton\rm}
\newcommand{\integ}{\it INTEGRAL\rm}

\newcommand{\chandra}{\it Chandra\rm}
\newcommand{\swift}{\it Swift\rm}

\newcommand{\vlt}{\it VLT\rm}
\newcommand{\hst}{\it HST\rm}
\newcommand{\keck}{\it Keck\rm}
\newcommand{\hess}{\it HESS\rm}
\newcommand{\simx}{\it Simbol--X\rm}

\begin{document}

\begin{frontmatter}

\title{Soft gamma-ray constraints \\ on a bright flare from \\ the Galactic Center \\ supermassive black hole}

\author{G. Trap$^{1,2}$, A. Goldwurm$^{1,2}$, R. Terrier$^{2}$, K. Dodds-Eden$^{3}$,}
\author{S. Gillessen$^{3}$, R. Genzel$^{3}$, E. Pantin$^{2,4}$, P.O. Lagage$^{2,4}$,}
\author{P. Ferrando$^{1,2}$, G. B\'elanger$^{5}$, D. Porquet$^{6}$, N. Grosso$^{6}$,}
\author{F. Yusef-Zadeh$^{7}$, F. Melia$^{8}$}

\thanks[1]{Service d'Astrophysique (SAp) / IRFU / DSM / CEA Saclay -- B\^at. 709, 91191 Gif-sur-Yvette Cedex, France}
\thanks[2]{AstroParticule \& Cosmologie (APC) / Universit\'e Paris VII / CNRS / CEA / Observatoire de Paris -- B\^at. Condorcet, 10, rue Alice Domon et L\'eonie Duquet, 75205 Paris Cedex 13, France, \texttt{trap@apc.univ-paris7.fr}}
\thanks[3]{Max Planck Institut f\"ur Extraterretrische Physik (MPE) -- 85748 Garching, Germany}
\thanks[4]{Astrophysique Interactions Multi-\'echelles (AIM) / Universit\'e Paris VII / CEA / CNRS -- B\^at. 709, 91191 Gif-sur-Yvette Cedex, France}
\thanks[5]{European Space Agency (ESA) / ESAC -- P.O. Box 78, Villanueva de la Canada, 28691 Madrid, Spain}
\thanks[6]{Observatoire astronomique de Strasbourg / Universit\'e de Strasbourg / CNRS / INSU -- 11, rue de l'Universit\'e, 67000 Strasbourg, France}
\thanks[7]{Department of Physics and Astronomy / Northwestern University, Evanston, Illinois 60208, USA}
\thanks[8]{Department of Physics and Steward Observatory / The University of Arizona, Tucson, Arizona 85721, USA}

\begin{abstract}

Sagittarius\,A$^\star$ (\sgra) is the supermassive black hole residing at the center of the Milky Way. It has been the main target of an extensive multiwavelength campaign we carried out in April 2007. Herein, we report the detection of a bright flare from the vicinity of the horizon, observed simultaneously in X-rays (\xmm/EPIC) and near infrared (\vlt/NACO) on April 4$^{\rm th}$ for 1--2~h. For the first time, such an event also benefitted
from a soft $\gamma$-rays (\integ/ISGRI) and mid infrared (\vlt/VISIR) coverage, which enabled us to derive upper limits at both ends of the flare spectral energy distribution (SED). We discuss the physical
implications of the contemporaneous light curves as well as the SED, in terms of synchrotron, synchrotron self-Compton and external Compton emission processes.

\end{abstract}


\end{frontmatter}

\parindent=0.5 cm


\section{Introduction}

From the discovery of a compact radio source, \sgra, at the Galactic Center (GC) in 1974 \citep{balick74} to the near infrared (NIR) tracking of stars in Keplerian motion around \sgra~three decades later \citep{schodel02,ghez03}, the evidence for a $\sim$$4\times10^{6}$~$M_\odot$ black hole with very slow proper motion at the dynamical center of our galaxy \citep{reid08} gradually piled up \citep[see][for a general review and references therein]{melia07}. 

Yet, the long quest for the high energy emission pertaining to the black hole has only been achieved recently. \sgra~was resolved as a notably dim ($2.4\times10^{33}$ \ergpersec, 2--10~keV) and slightly extended (1.4$''$) point source with the \chandra~satellite in 1999 \citep{baganoff03a}. One year later, the same instrument witnessed the source exhibiting an X-ray flare for $\sim$3 h \citep{baganoff01}. A $\sim$10~min long substructure within the light curve of the eruption and light time travel arguments imply that this event took place close to the  event horizon ($<15~R_{\rm S}$).
Many other detections of X-ray flares followed, either with \xmm~or \chandra~\citep[see e.g.][]{goldwurm03a,baganoff03b,porquet03,belanger05}, and established that the duty cycle of the black hole is nearly one X-ray flare per day. 
The origin of these events is still unclear, in spite of all the efforts aimed at their monitoring in different energy ranges. 
In 2003, NIR flares from \sgra~were indeed discovered with the \vlt~\citep{genzel03}, and later confirmed by the \keck~\citep{ghez04} and the \hst~\citep{yusef-zadeh06a}. 
They occur more frequently than the X-ray ones (around four per day) and have been observed in many NIR atmospheric pass bands (H, K, L, M). Each new infrared flare has generally induced either spatial \citep{clenet05}, spectral \citep{eisenhauer05,ghez05,gillessen06,krabbe06,hornstein07}, polarimetric \citep{eckart06b,meyer06,meyer07,trippe07}, or timing studies \citep{meyer08,do09}.
Numerous multiwavelength campaigns showed that an X-ray flare always comes along with a  simultaneous NIR one\footnote{The converse is not true, some NIR flares have no X-ray counterpart \citep{hornstein07}.} \citep{eckart04,eckart06a,eckart08a,yusef-zadeh06a,hornstein07}, and maybe a delayed submm one \citep{marrone08,eckart08b,yusef-zadeh08} caused by plasmon expansion \citep{liu04,yusef-zadeh06b}. 

Above 6~$\mu$m, in the mid infrared (MIR), no detection of \sgra~has been reported so far. Recent upper limits on the black hole flux at 8.6~$\mu$m were set by the \vlt/VISIR instrument during low level NIR variability by \citet{schodel07}, who argued that a detection would be reachable in case of a strong NIR flare. 

Above 20 keV, repeated surveys of the heart of the Milky Way in soft $\gamma$-rays with the \integ~satellite unveiled a persistent pointlike source compatible with \sgra~location (within the 1$'$ error radius), \igr~\citep{belanger04,belanger06}. The nature of the source is still uncertain, and a possible association with the supermassive black hole remains conceivable. Given the limited angular resolution of the soft $\gamma$-ray telescope \integ/IBIS/ISGRI ($\sim$12$'$ FWHM), the best way to unequivocally identify the mysterious \igr~with \sgra~is the detection of correlated variability between soft $\gamma$-rays and other wavelengths.

To tackle the above puzzles and investigate the correlated X-ray/NIR variability of \sgra~in more details, a coordinated multiwavelength campaign on the GC was conducted in spring 2007. It involved in particular the \xmm~and \integ~satellites for the high energies, as well as the \vlt/ NACO and \vlt/VISIR ground instruments to cover the NIR and MIR part of the spectrum, respectively. Their results are presented in Sect.~\ref{obs} and interpreted in Sect.~\ref{SED}. Note that the X-ray and infrared findings have already been published by \citet{porquet08} and \citet{dodds09}, respectively. 

We will not discuss here the short term variability of \sgra~in April 2007 at cm, mm, and submm wavelengths, which will be reported in another article, along with NIR results obtained by the Hubble Space Telescope \citep{yusef-zadeh09}\footnote{For further discussion of the past variability of \sgra~in cm and mm bands, see for example \citet{zhao01,herrnstein04} and \citet{tsuboi99,zhao03}, respectively.}. 

Throughout this paper we adopt a GC distance of 8 kpc \citep{reid93} and a black hole mass $M_{\bullet}=4\times10^{6}$~$M_\odot$ \citep{ghez08}, for which the Schwarzschild radius is $R_{\rm S}=1.2\times10^{12}$~cm.  


 \begin{figure*} 

   \centering
   \includegraphics[trim=1cm 2cm 1cm 1.5cm, clip=true, width=14cm]{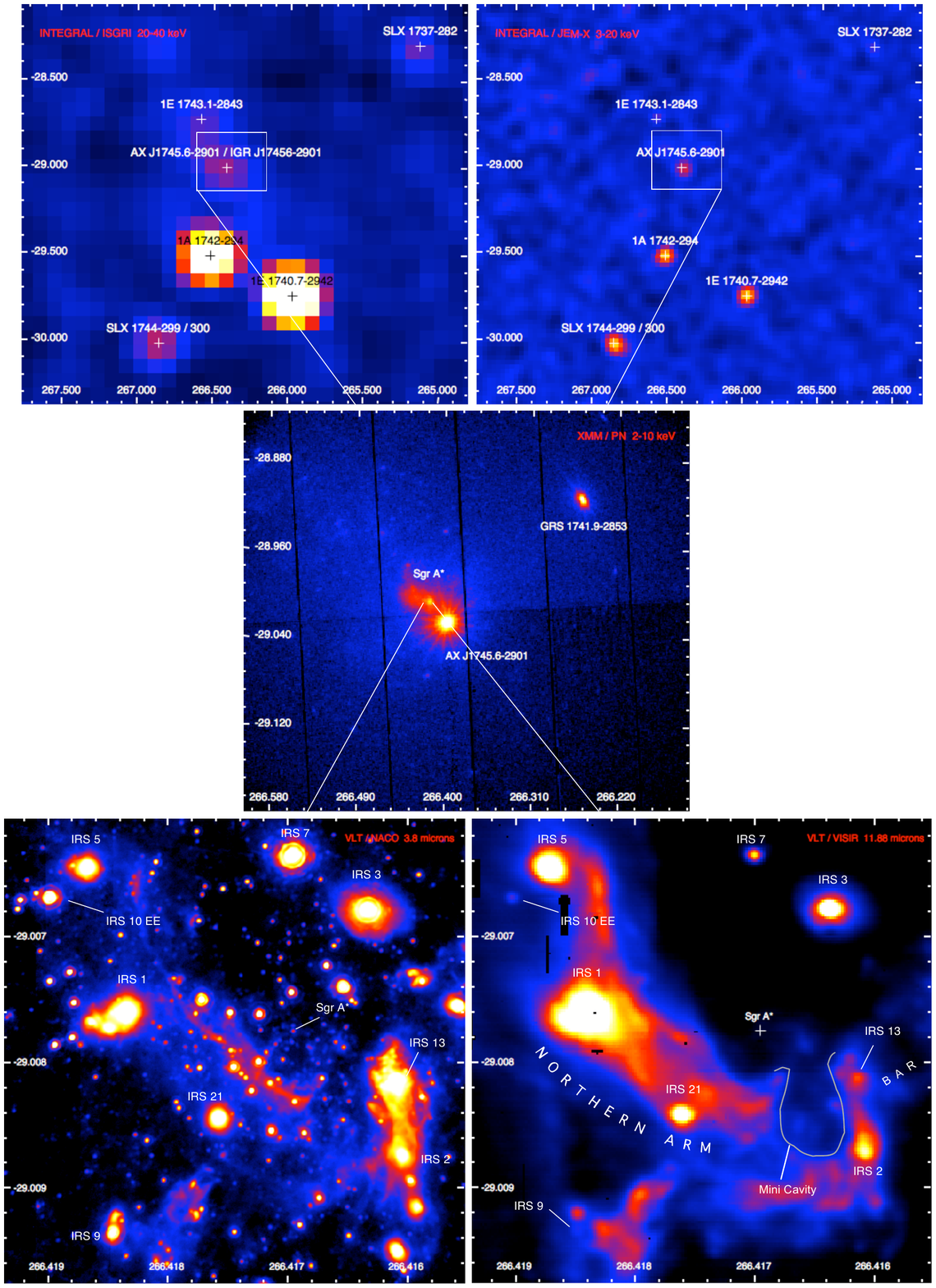}
   \caption{Multiwavelength and multiscale views of the Galactic Center in April 2007 in R.A. ($^\circ$) horizontally and Dec. ($^\circ$) vertically (North and East point towards the top and the left, respectively). \emph{Top, left}: INTEGRAL/ISGRI mosaic  in the 20--40~keV band. The Galactic plane runs from upper left to bottom right. \emph{Top, right}: INTEGRAL/JEM-X 1 mosaic  in the 3--20~keV band. \emph{Middle}: XMM/PN image in the 2--10~keV band. \emph{Bottom, left}: VLT/NACO image at the peak of the flaring period at 3.8~$\mu$m. \emph{Bottom, right}: VLT/VISIR average image of the 3$^{\rm rd}$/4$^{\rm th}$ April night at 11.88~$\mu$m. The three dusty arms swirling around \sgra~compose the so-called ``mini-spiral''.}
   \label{ima}
\end{figure*}

\section{Observations \& results}
\label{obs}

\subsection{X-rays} 
\label{X-rays}

The \xmm~satellite \citep{jansen01} was pointed towards the GC during $\sim$2.5 consecutive revolutions, from March 30$^{\rm th}$ to April 4$^{\rm th}$ 2007. The data of the EPIC/PN \citep{struder01} and EPIC/MOS1--2 cameras \citep{turner01} were processed and analyzed through the procedure described in \citet{porquet08}. 

We produced an image of the last revolution of the campaign (rev-1340, 97.6~ks exposure), cleaned for out of time events in the 2--10~keV band (see Fig.~\ref{ima}, middle). Two GC transient X-ray binaries and bursters, active at the time, stand out prominently: \grs~\citep{trap09} and \ax~\citep{grosso08}. \sgra~is clearly apparent in the middle since this observation contains several flares from the vicinity of the black hole, enhancing its average luminosity.

\begin{figure*} 
   \centering
   \includegraphics[trim=0cm 1.5cm 0cm 0cm, clip=true, width=14cm]{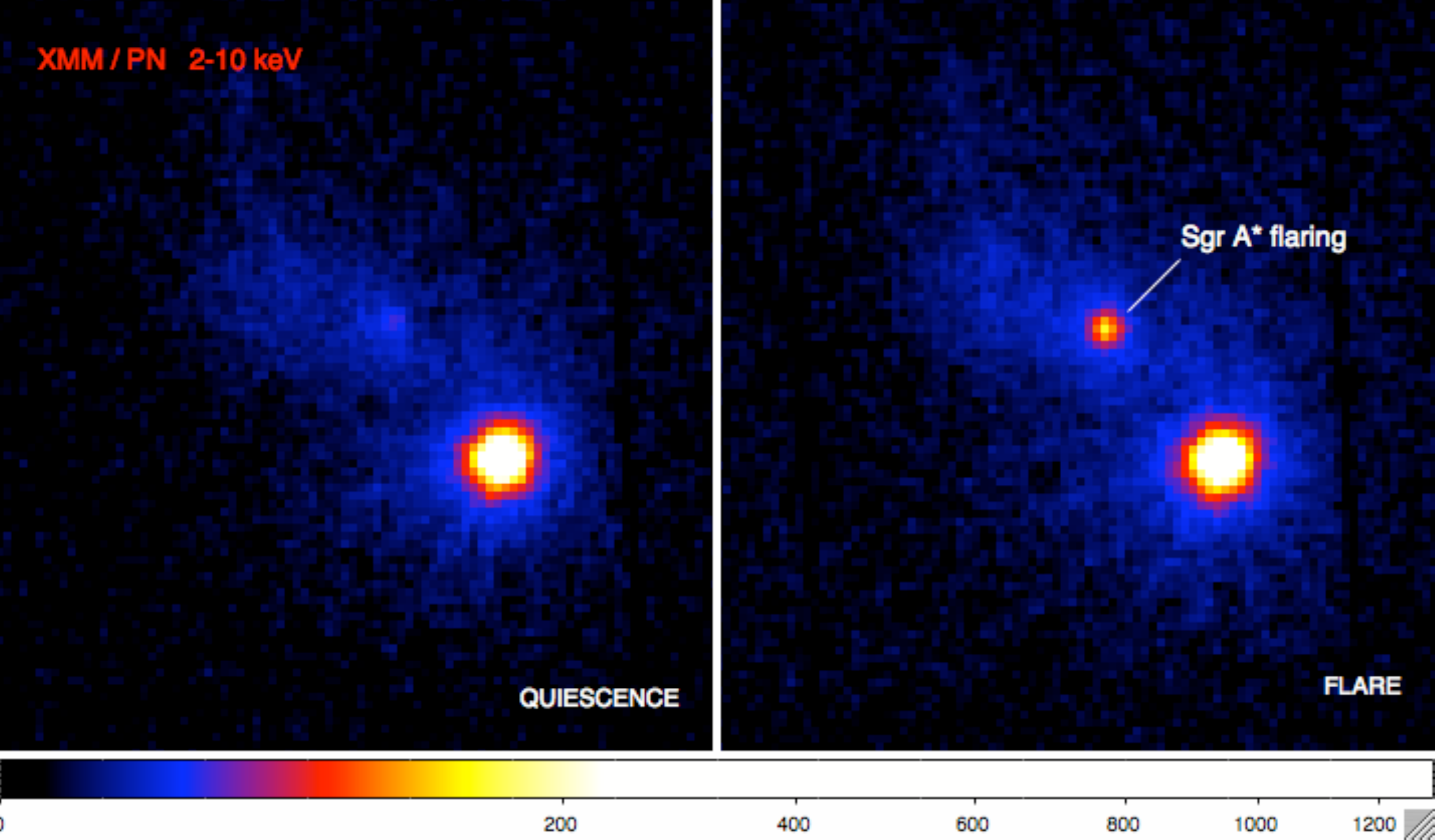}
   \caption{{\it Left}: X-ray image of the quiescent phase during 50~min before the flare. {\it Right}: 50~min image during the flare.}
   \label{imax}
\end{figure*}

Indeed, on April 4$^{\rm th}$, a high level of flaring activity from \sgra~was caught. A bright flare (Fig.~\ref{imax})---the second brightest ever recorded ($\sim$100 times the quiescent level) in the X-ray band (2--10 keV)---was rapidly followed by three moderate ones. The bright event lasted $\sim$1 h; its PN light curve has a rather symmetrical morphology and no apparent substructures (Fig.~\ref{lc}). Note that the 10$''$ radius area used to extract this light curve not only contains \sgra~but other X-ray sources as well: a pulsar wind nebula candidate, \pwin~\\
\citep{wang06}, the star cluster IRS\,13, and diffuse emission \citep{baganoff03a}, which all, however, provide a constant contribution.

 From a spectral point of view, this outburst was rather soft. The best fit to the data with an absorbed power-law model, including dust scattering, yields the following parameters: a spectral photon index $\Gamma = 2.3 \pm 0.3$ ($N(E)\propto E^{-\Gamma}$) and a column density $N_{\rm H} = 12\pm2 \times 10^{22}$ \cm. In Sect.~\ref{SED}, we will use an equivalent definition of the spectral index, $\beta_{\rm X}$, easier to compare to other multiwavelength spectra: $\beta_{\rm X}=-\Gamma+2=-0.3\pm0.3$ with  
$\nu F_{\nu}^{\rm X} \propto \nu^{\beta_{\rm X}}$.
 The unabsorbed mean flux of the flare was $16\pm3 \times 10^{-12}$~\ergperseccm~(2--10~keV), which translates to a luminosity of $2.4\pm4 \times 10^{35}$ \ergpersec~at the GC distance. Albeit luminous relative to previously observed X-ray flares, it was still $\sim$9 orders of magnitude below the Eddington luminosity for a supermassive black hole of this kind. 

As pointed out by \citet{porquet08}, this rapid train of flares in just a few hours challenges disruption mechanisms of the accretion flow as the origin of the outbursts, since they rely on temporary storage of mass/energy. This energy should indeed be released at once during the outburst, with a radiation efficiency of a few percent. But, the weak accretion rate of the black hole seems insufficient ($\sim$$10^{16-17}$ g\,s$^{-1}$, \citet{melia07}) to accumulate the required energy on such short timescales. In contrast, scenarios based on the stochastic infall and tidal disruption of gas clumps \citep{tagger06,falanga07,falanga08} or small bodies \citep{cadez08} do not encounter this issue.  

\subsection{Near infrared}
\label{NIR}

The \vlt/NACO (NAOS+CONICA) set of instruments \citep{lenzen03,rousset03} installed on the ESO/\vlt~unit telescope Yepun (UT4) at Paranal, Chile, observed the Galactic nucleus every nights from April 1$^{\rm st}$ to April 6$^{\rm th}$ in multiple NIR bands: L' (3.8~$\mu$m), K$_{\rm S}$ (2.1~$\mu$m), and H (1.6~$\mu$m). The details of the data reduction and analysis are presented in \citet{dodds09}. 

\begin{figure*} 
   \centering
   \includegraphics[trim=0cm 1.5cm 0cm 0cm, clip=true, width=14cm]{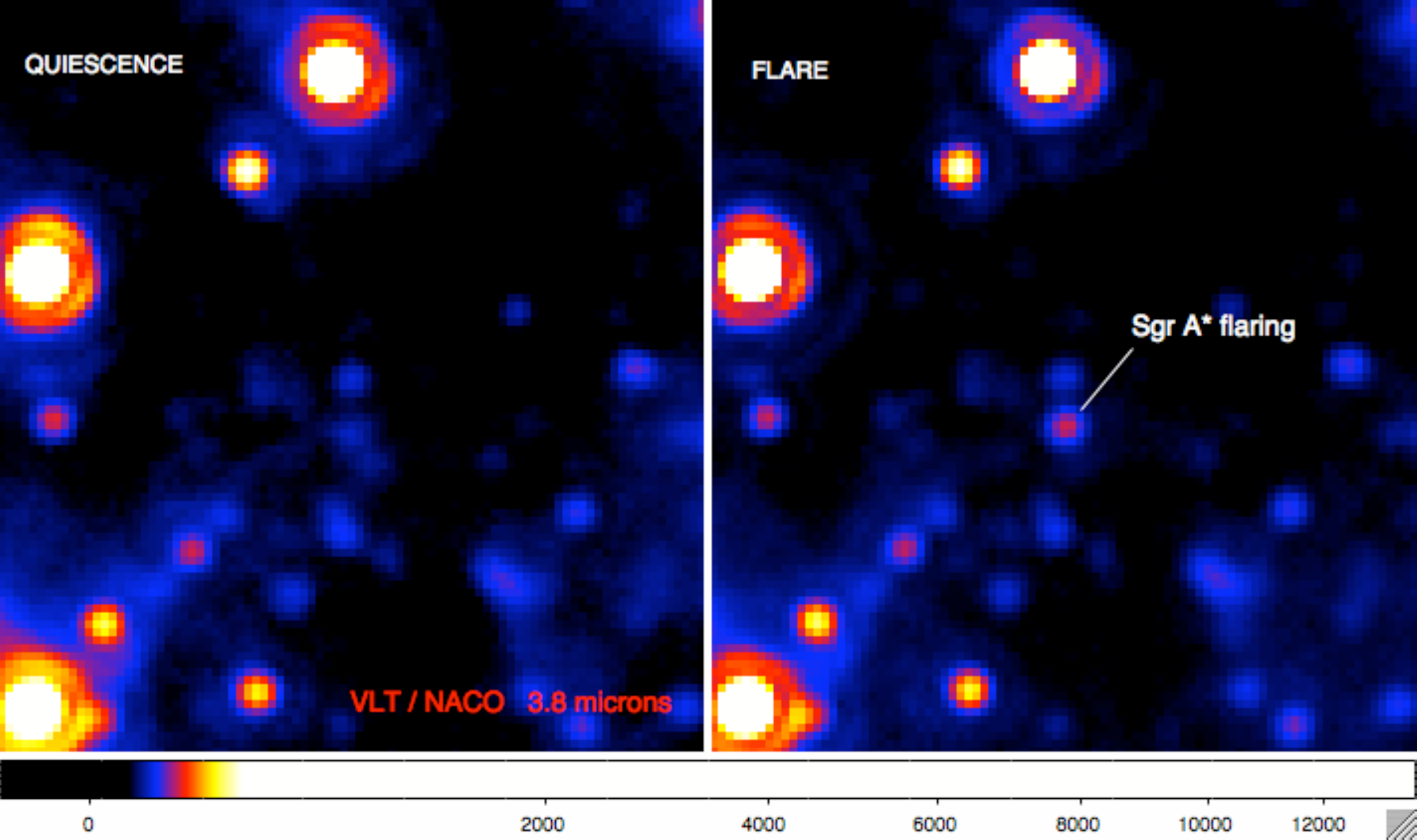}
   \caption{{\it Left}: NIR image of the quiescent phase during six minutes before the flare. {\it Right}: Six minutes image at the apogee of the flare.}
     \label{imanir}
\end{figure*}

In particular, from 5:00 to 7:00 (UT) on April 4$^{\rm th}$, NACO followed the strong X-ray flare mentionned in Sect.~\ref{X-rays} in the L' band. We constructed an image of the GC during this flare period (Fig.~\ref{ima}, bottom, left, and Fig.~\ref{imanir}), in which \sgra~is confused with the star S\,17 and a small cloud of dust \citep{clenet05}. On Fig.~\ref{lc}, we display the light curve of the flare. Substructures on a timescale of $\sim$20~min within the light curve are evident. The shortest variation in the light curve ($\Delta t\sim1$~min) constrains the emitting zone to a size no bigger than $c/\Delta t \sim 1.5~R_{\rm S}$. This is the first time that such features are visible in the L' band. Regarding the total duration of the eruption, it lasted distinctively longer in the NIR than in X-rays ($\sim$2~h vs. $\sim$1~h), even though the X-ray background could hide the rising and decaying flanks of the flare. Note that the X-ray/NIR flare detected by \chandra/\keck~on July 17$^{\rm th}$ 2006 \citep{hornstein07} also indicated a longer NIR duration, even if \chandra~has a smaller X-ray background than \xmm~thanks to its better point spread function (PSF). As for the peak of the flare, there is no time lag bigger than $\sim$3~min between NIR wavelengths and X-rays.    

Assuming an extinction $A_{\rm L}= 1.8$~mag, the NIR flare peaked at $\sim$30~mJy (dereddened), which makes it one of the most powerful NIR flare ever captured, and definitely the brightest one detected simultaneously in X-rays. To allow comparison with other wavelengths, we also computed the background subtracted, extinction corrected, mean flux of the flare over the period of MIR observations (see Sect.~\ref{MIR}): $19.1\pm3.6$~mJy at 3.8~$\mu$m. No direct NIR spectral information are available for this flare.

\subsection{Mid infrared}
\label{MIR}

VISIR, the \vlt~Imager and Spectrometer for the mid Infrared mounted on the ESO/\vlt~unit telescope Melipal (UT3) \citep{lagage04,pantin05}, pointed the GC from 2007-04-04 5:29:00 to 2007-04-04 10:34:00 (UT) as part of a guaranteed time program. Data were acquired with the imager and {\tt PAH2\_2} filter on, at $11.88\pm0.37~\mu$m in the atmospheric window N. The Small Field mode ({\tt \small SF}) was employed, which resulted in a field of view of $256 \times 256$ pixels ($19.2''\times19.2''$), each pixel corresponding to $0.075''\times0.075''$. Data reduction and analysis techniques are given in \citet{dodds09}. 

\begin{figure} 
 \centering
 \includegraphics[trim=0cm 7cm 0cm 1cm, clip=true, width=14cm]{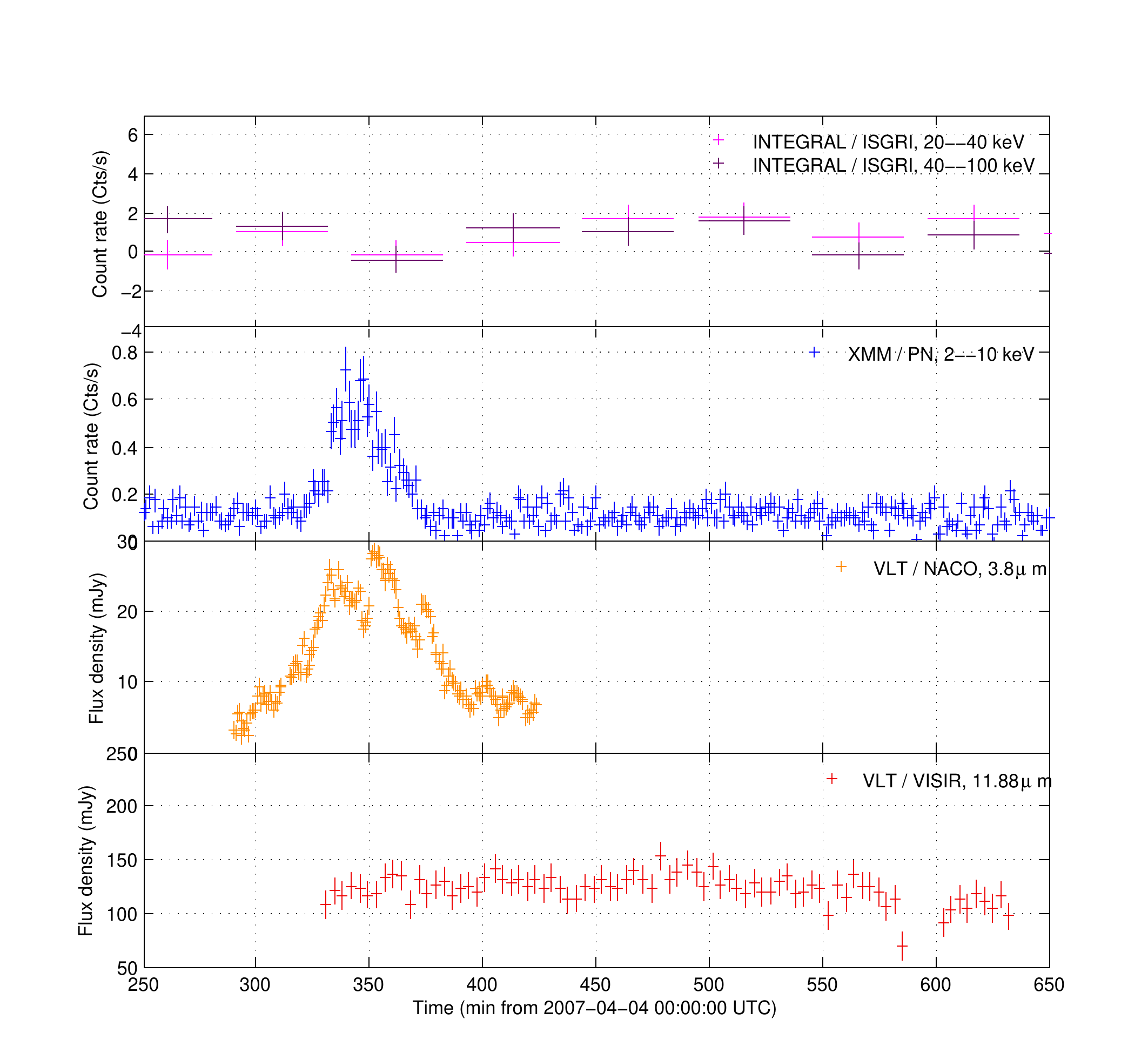}
 \caption{From top to bottom, light curves of \igr~+ \ax, \sgra~+ \pwin~+ IRS\,13 + diffuse emission, \sgra + S\,17, and diffuse emission.}
\label{lc}
\end{figure}

No point source at the position of \sgra~is detected in either the individual images or the collapsed image of the entire night (Fig.~\ref{ima}, bottom, right). We also performed a Lucy-Richardson deconvolution with HD\,102461 as point spread function without success. The flux from a box of $0.375''\times0.375''$ centered on the position of \sgra~is fairly constant (see the light curve on Fig.~\ref{lc}) with an average value of $123\pm6$ mJy (not dereddened). This flux may be attributed to the faint and diffuse dust ridge on which \sgra~lies. Our measured value is consistent with previous VISIR observations \citep{eckart04,schodel07} and other instruments before \citep{stolovy96,cotera99,morris01}.

However, for the first time, the measurements presented here were concurrent with a bright X-ray/NIR flare from \sgra~as shown in Sect.~\ref{X-rays} and \ref{NIR}. 
We estimate that \sgra~could not have been brighter than $\sim$12~mJy at 11.88~$\mu$m (3$\sigma$, not dereddened). This value is compatible with VISIR empirical sensitivity at this wavelength: 7~mJy/10$\sigma$/1~h (median value for different atmospheric conditions). We note also that similar constraints were obtained with VISIR during NIR variability by \citet{gillessen06}, \citet{schodel07}, and \citet{haubois08}.
 
The value of the extinction correction, $A_{\rm \lambda}$, in the MIR depends critically on the strength and shape of the silicate absorption feature at $\sim$10~$\mu$m. In the literature, $A_{\rm \lambda}$ is usually given as ratios relative to $A_{\rm V}$ or $A_{\rm K}$, so we use $A_{\rm K} = 2.8$~mag ($A_{\rm V} = 25$~mag) to ensure consistency across our multiwavelength observations. The closest extinction measurement to $\lambda = 11.88$~$\mu$m was made by \citet{lutz99} for a wavelength of $\sim$12.4~$\mu$m. We consider the recent theoretical model of \citet{chiar06} for the extinction profile in the silicate region to allow us to extrapolate the value measured at 12.4~$\mu$m to 11.88~$\mu$m. When normalized to the \citet{lutz99} values, the model predicts $A_{\rm 11.88\,\mu m}=1.7 \pm 0.2$~mag and hence the dereddened 3$\sigma$ upper limit on the MIR emission of \sgra~during the flare is $\sim$57~mJy. \citet{schodel07} found an upper limit of $\sim$22~mJy at 8.6~$\mu$m during quiescence of \sgra~and predicted a {\it bright} flare was likely to come out of the noise in the MIR. Yet, using the dereddening used here, their limit would be a bit higher, $\sim$32~mJy, and, since we did not detect this strong X-ray/NIR flare, we speculate that no flaring counterpart of \sgra~will be detected with the current settings of VISIR.

\subsection{Gamma-rays}

The \integ~satellite \citep{winkler03} monitored the GC in parallel to the other instruments in April 2007, for a total effective exposure of $\sim$212 ks for IBIS/ISGRI (20--100 keV) \citep{ubertini03,lebrun03} and $\sim$46 ks for JEM-X 1 (3--20 keV) \citep{lund03}\footnote{Compared to IBIS, JEM-X field of view is narrower, so given the rectangular dithering pattern used ($5 \times 5$ pointings = 24 off-source + 1 on-source), the GC was invisible to JEM-X most of the time, which explains the discrepancy in the exposures.} at \sgra~position. 
Measurements were spread over two consecutive revolutions, 545 and 546, from 2007-04-01 12:58:00 to 2007-04-02 21:32:34 and 2007-04-03 11:48:14 to 2007-04-04 20:26:59 (UT), respectively. In total, data of 74 individual pointings (science windows, ScWs) were acquired,  lasting $\sim$2930 s each. The whole dataset was reduced with OSA 7.0, the Offline Science Analysis Software, distributed by the \integ~Science Data Center (ISDC) \citep{courvoisier03}, with algorithms described in \citet{goldwurm03b} for IBIS/ISGRI and \citet{westergaard03} for JEM-X. 

To search for a counterpart above 20 keV of the aforementioned flare, we selected the two consecutive ScWs of ISGRI that covered the flare time interval, 054600220010 and 054600230010, and created a combined mosaic of the individual images, in two energy bands: 20--40 and 40--100~keV.   
None of these mosaics contained a distinctive source at the position of the black hole. Hence, no high energy counterpart of the flare was found. By considering the variance in \sgra~pixel, we derive 3$\sigma$ upper limits on the flare of 1.17 and 1.11 \ctspersec~in the 20--40 and 40--100 keV bands, respectively. Assuming a power-law spectral shape of index $\Gamma=2.3$ (Sect.~\ref{X-rays}), these rates convert to flux limits of 5.76 and $11.1\times10^{-11}$ \ergperseccm, respectively.      

Regarding JEM-X 1 data in the 3--20 keV band, we report similar results. There was no detection in the combined mosaic, and given a flare duration of $\sim$3000 s, the sensitivity curves of JEM-X\footnote{See JEM-X user manual: \\ \tt \small http://isdc.unige.ch/Soft/download/osa/osa\_doc/prod/osa\_um\_ibis-7.0.pdf} provide 5$\sigma$ upperlimits of 10 and $7\times10^{-11}$ \ergperseccm~in the 3--10 and 10--25 keV energy ranges, respectively. 
 
On Fig.~\ref{lc} (top panel), we plotted the ISGRI light curves of the pixel at the position of \sgra, built with individual ScW. It is noteworthy that no source was significantly detected in any individual exposure. 

In contrast to individual ScWs, the total ISGRI and JEM-X 1 mosaics of the observation dataset both reveal a significant excess at the position of \sgra~(see Fig.~\ref{ima}, top, left, and right). The significances of these signals are both 13.7$\sigma$. In view of the \xmm~image (Fig.~\ref{ima}, middle), the transient neutron star low-mass X-ray binary \ax, located just 1.5$'$ from \sgra~in projection, was markedly the dominant source of the region. Given JEM-X angular resolution of $\sim$3$'$ (FWHM), we can safely associate its 3--20 keV excess with the binary. In the ISGRI mosaic (Fig.~\ref{ima}, top, left), the PSF is $\sim$13$'$ (FWHM), and so does not allow us to disentangle \ax~from \igr, the persistent hard X-ray source discovered by \integ/ISGRI\footnote{Notice that \igr~has not been significantly detected with JEM-X yet.}~\citep{belanger04,belanger06}. To assess the contribution of the transient binary to the ISGRI signal though, we compared the April 2007 20--40 keV mosaic with another equivalent GC map, constructed with data spanning four months, from August to November 2006. During this latter period, we know for sure that the transient binary was in quiescence and undetected at high energies, thanks to a regular \swift/XRT monitoring of the GC \citep{degenaar09}. So, the total count rate of $0.86\pm0.03$ \ctspersec~we measured in the central pixel of the excess in the 2006 mosaic, can be entirely attributed to \igr. In April 2007, we found that the total count rate increased to $0.97\pm0.07$ \ctspersec, so that, presuming \igr~remained constant, the photons from the 20--40 keV excess visible in Fig.~\ref{ima} (top, left) came at $\sim$90\% from \igr~and $\sim$10\% from \ax.   

This is the first time \integ~was gazing the GC during a period of known flaring activity from \sgra. Previous X-ray/$\gamma$-ray coordinated campaigns in 2004 were, indeed, inconclusive, since the X-ray flares detected then by \xmm~occured at times when \integ~was crossing the radiation belts with all its intruments in standby mode \citep{belanger06}.  

As indicated above, we did not identify any $\gamma$-ray counterpart of the intense X-ray flare from April 4$^{\rm th}$. This proves once again that \sgra~does not release the bulk of its emission in soft $\gamma$-rays \citep{goldwurm94}. This result is also somewhat reminescent of the 2005 \chandra/\hess~joint campaign, which demonstrated that the TeV source of the GC, \hesssource, stayed still during an X-ray outburst seen by \chandra~\citep{hinton07a,aharonian08}. 

On Fig.~\ref{sed}, we display the broad band quiescent spectral energy distribution (SED) of \sgra~in dark gray. We also overplot in blue the spectral information on the 2007 April 4$^{\rm th}$ flare. 
By extrapolating the X-ray power-law, one expects fluxes of 3.9 and $4.1\times10^{-12}$ \ergperseccm~in the 20--40 and 40--100 keV, respectively. These expected values are roughly one order of magnitude below the 3$\sigma$ constraints worked out above, which suggest that the next generation of hard X-ray focusing instruments, like \simx~\citep{ferrando08}, will be able to extend spectral measurements on a flare of the GC supermassive black hole to above 20 keV \citep{goldwurm08}.  

Concerning \igr, it was relatively improbable to find it flare up in April 2007, based on the long ISGRI exposures targeted at the source in 2003 and 2004. These did not reveal any sign of variability on any timescale \citep{belanger06}, despite a temporal artefact in the early light curves of \igr~\citep{belanger04}, that was later attibuted to a poor correction of the background \citep{belanger06}.
Note, however, that variability on a single ScW duration basis cannot really be excluded, since this time interval is too short to convincingly detect the source \igr. 

The provenance of \igr~thus remains enigmatic. We showed that the activity of the luminous transient binary \ax~did not amount to more than $\sim$10\% of the total 20--40 keV flux of \igr, contrary to what was alluded to in \citet{revnivtsev04}. The absence of variability and the fact that the flux of \igr~is two orders of magnitude above the quiescent emission of \sgra~as measured by \chandra, supports the idea that the hard X-ray photons visible in \integ's mosaics are unlikely to be produced in the inner region of the accretion/ejection flow around the black hole. Instead, these photons should arise from a diffuse, and yet compact (a few arcminutes), zone, or maybe result from the sum of unresolved hard X-ray point sources \citep{revnivtsev06}. 
A possible connection between \igr~and \hesssource~is another option. \citet{hinton07b} put forward that $\sim$10--100 TeV electrons permeating the inner 20~pc may be responsible for the combined \xmm/\integ~spectrum of the central 8$'$ radius region \citep{belanger06} via synchrotron emission, as well as \hesssource~through inverse Compton (IC) processes. These authors favor the pulsar wind nebula candidate \pwin~\citep{wang06} as the X-ray counterpart of \hesssource, though. In their scenario, the TeV photons come about in the compact nebula, just 0.3~pc from \sgra, by the IC boosting of ambient photons by relativistic electrons originating from the pulsar. Nevertheless, \igr~does not fit within this frame, as its flux is too high to be the simple hard X-ray extension of \pwin~soft X-ray flux as determined by \chandra~\citep{wang06}. The increased angular resolution and sensitivity in the hard X-ray range of the next generation of instruments will also help address the question of \igr~true nature.

\section{Radiative processes}
\label{SED}
 
The extremely low quiescent luminosity of \sgra~(10 orders of magnitude below the Eddington luminosity) is a long standing puzzle, that has stimulated numerous theoretical studies based on a combination of a low accretion rate, a radiatively inefficient accretion flow and outflows ejecting out the matter that just flowed in \citep[See][for a review and references therein]{melia01}. We will not discuss here the quiescent state and rather concentrate on the flaring emission, for which many models have also been proposed: jet models \citep{markoff01}, ADAF like models \citep{yuan03,yuan04} and accretion/stochastic acceleration models \citep{liu02,liu04,liu06a,liu06b}. Each of them usually invokes either synchrotron self-Compton (SSC), external Compton (EC) or synchrotron broken power-laws (SB) processes as radiation mechanisms. In the subsequent discussion, we will explore these different possibilities for the April 4$^{\rm th}$ event, with no {\it a priori} assumption about the true nature of the engine behind the flare. We will examine the case of power-law distributions of electrons and highlight the natural synchrotron self-Compton component of each model.

\subsection{Synchrotron self-Compton}
\label{ssc}

NIR flares are traditionally thought to arise from synchrotron emission since they are highly polarized \citep{eisenhauer05,gillessen06,krabbe06,hornstein07} and have power-law spectral shapes \citep{eckart06b,meyer06,meyer07,trippe07}. Here, we could not obtain a direct NIR spectrum of the flare, but we do have a stringent MIR upper limit. Hence, supposing a power-law spectral shape from MIR to NIR, $\nu F_{\nu}^{\rm IR} \propto \nu^{\beta_{\rm IR}}$, we have $\beta_{\rm IR}>0.04$. This is consistent with the index $\beta_{\rm NIR}=0.4$ published in previous NIR studies \citep{genzel03,gillessen06,hornstein07}. 
The submm bump of \sgra~quiescent SED has a slope $\beta_{\rm submm}^{\rm thick}>0$ below $\sim$$10^{12}$~Hz, which is thought to arise from an optically thick regime, and a slope $\beta_{\rm submm}^{\rm thin}<0$ above $\sim$$10^{12}$~Hz, presumably coming from an optically thin regime, judging from polarization measurements \citep{aitken00}. The fact that $\beta_{\rm IR}>0$ around $10^{13}$~Hz, shows that the NIR flare did not come from a global shift upward of the submm bump, but from a distinct population of particles creating a new rising hump in the SED in the IR band (see Fig.~\ref{sed}).

In an SSC model for the flares, the NIR photons are produced by a momentarily accelerated population of electrons radiating in the NIR band via a synchrotron process.  In the following we will use the simple parametrization of \citet{kraw04} in which a spherical homogeneous source of synchrotron radiation with a radius $R$ and a volumic electron density $n_{\rm e}$, pervaded by a magnetic field $B$, has a power-law energy distribution:

$$ n(\gamma) \propto \gamma^{-p}~~~{\rm for}~~~\gamma_{\rm min} < \gamma < \gamma_{\rm max}\;.$$

We set $p=2$ in what follows\footnote{Such an index could be the natural consequence of a Fermi II acceleration process for instance.}. The electron density is thus determined by the normalization factor, $n_0$, of the power-law distribution by: 
 
 $$n_{\rm e}=\int^{\gamma_{\rm max}}_{\gamma_{\rm min}} n(\gamma)\, d\gamma =  \int^{\gamma_{\rm max}}_{\gamma_{\rm min}} n_0 \gamma^{-2}\, d\gamma = -n_0 (\gamma_{\rm max}^{-1}-\gamma_{\rm min}^{-1}) \;.$$
 
The energy density of the electrons, $w_{\rm e}$, used by \citet{kraw04} as normalization is also linked to $n_0$ through:

$$w_{\rm e}=\int^{\gamma_{\rm max}}_{\gamma_{\rm min}} \gamma m_{\rm e} c^2 n(\gamma)\, d\gamma=n_0 m_{\rm e} c^2 \ln{\gamma_{\rm max} \over \gamma_{\rm min}} \;.$$

The resulting synchrotron photon spectrum is optically thin and has a power-law shape \citep{rybicki79}:

$$\nu F_{\nu}^{\rm sync} \propto n_{\rm e} R^3 B^{(1+p)/2}  \nu^{(3-p)/2}~~~{\rm for}~~~\nu_{\rm min}^{\rm sync} < \nu < \nu_{\rm max}^{\rm sync}\;,$$

with  $\nu_{\rm min(max)}^{\rm sync}\propto \gamma_{\rm min(max)}^2 \nu_{g}$, where $\nu_{\rm g}=\frac{eB}{2\pi m_{\rm e}c}$ is the gyration frequency and $m_{\rm e}$ the mass of the electron. 
Below $\nu_{\rm min}^{\rm sync} $, the photon spectrum has a power-law shape, $\nu F_{\nu}^{\rm sync} \propto  \nu^{4\over3}$, due to the lowest energetic electrons which have a Lorentz factor $\gamma_{\rm min}$, and above $\nu_{\rm max}^{\rm sync} $ it has an exponential cut-off.

In this SSC scheme, the X-ray flare is provoked by the inverse Compton boosting of the NIR flare photons by the same electrons that have just given rise to the NIR photons. The inverse Compton spectrum has the same morphology as the synchrotron one and scales like the Thomson optical depth of the sphere, $n_{\rm e} R \sigma_{\rm T}$, times $\nu F_{\nu}^{\rm sync}$: 

$$\nu F_{\nu}^{\rm ic} \propto n_{\rm e} R \sigma_{\rm T} \times \nu F_{\nu}^{\rm sync} \propto n_{\rm e}^2 R^4 B^{(1+p)/2}  \nu^{(3-p)/2}~~~{\rm for}~~~\nu_{\rm min}^{\rm ic}<\nu<\nu_{\rm max}^{\rm ic}\;,$$ 

where $\sigma_{\rm T}$ stands for the Thomson cross-section\footnote{$\sigma_{\rm T} = \frac{8}{3}\pi r_{\rm e}^2$, where $r_{\rm e}=e^2/(m_{\rm e} c^2)= 2.82 \times 10^{-13}$~cm is the classical radius of the electron.} and:
 
$$\nu_{\rm min(max)}^{\rm ic}\propto \gamma_{\rm min(max)}^2 \nu_{\rm min(max)}^{\rm sync}\;.$$ 

Hence, this model has six free parameters: $p$, $\gamma_{\rm min}$, $\gamma_{\rm max}$, $B$, $R$, and $n_{\rm e}$. We arbitrarily fix $p=2$, $\gamma_{\rm min}=1$, and assess the four other parameters by considering four observables: the two frequencies of the synchrotron, $\nu_{\rm max}^{\rm sync}$, and inverse Compton, $\nu_{\rm max}^{\rm ic}$, peaks and their two respective amplitudes, $\nu F_{\nu}^{\rm sync,max}$ and $\nu F_{\nu}^{\rm ic,max}$. Indeed,  on the one hand, $\gamma_{\rm max}$ is given by:  

\begin{equation}
\label{gamma}
\nu_{\rm max}^{\rm ic}/\nu_{\rm max}^{\rm sync}\simeq \gamma_{\rm max}^2\;,
\end{equation}

and $B$ by \citep{rybicki79}\footnote{This is an upper limit obtained for a pitch angle of $\pi/2$.}:

\begin{equation}
\label{B}
\nu_{\rm max}^{\rm sync}\simeq 2.8 \left( \frac{B}{1~{\rm G}}\right) \gamma_{\rm max}^2 \times 10^{6}~{\rm Hz}\;.
\end{equation}

On the other hand, $R$ and $n_{\rm e}$ can be deduced from the relations:

$$ \left\{ \begin{array}{ll} 
\displaystyle\frac{\nu F_{\nu}^{\rm ic,max}}{\nu F_{\nu}^{\rm sync,max}} & \propto n_{\rm e}R  \\
\nu F_{\nu}^{\rm sync,max} & \propto n_{\rm e}R^3   
\end{array} \right.\;.$$

Yet, our measurements do not provide us these four observables {\it per se}. We know that the X-ray spectral slope is softer than the IR one, which suggests that our \xmm~measurement at $\nu_{\rm X}\approx10^{18}$~Hz lies between the inverse Compton peak and the cut-off. Regarding the synchrotron peak, we only know it has a frequency  $\nu_{\rm max}^{\rm sync}>10^{14}$~Hz. We will presume that this peak occurs at $\sim$$10^{14}$~Hz, in order to keep the magnetic field, $B$, as low as possible. As a result, we do not have to introduce the Klein-Nishina cross sections since the upscattering of the seed photons statisfies the Thomson regime condition, $\gamma_{\rm max} h \nu_{\rm seed} \ll m_{\rm e}c^2$ (the transition to the Klein-Nishina regime is at $\sim$$10^{18}$~Hz). 

With the set of parameters listed in Tab.~\ref{table} we obtain a good fit of the flare SED, as displayed in orange on Fig.~\ref{sed}. However the values of $B$ and $n_{\rm e}$ necessary to accommodate the data are extremely high. First, synchrotron cooling of the radiating particles has a characteristic timescale:

$$\tau_{\rm cool} = \left(\frac{1}{6 \pi} \frac{\sigma_{\rm T} B^2}{m_{\rm e} c} \gamma (1-\gamma^{-2})    \right)^{-1}
=1.3  \left(\frac{\nu}{1~{\rm Hz}}\right)^{-\frac{1}{2}} \left(\frac{B}{1~{\rm G}}\right)^{-\frac{3}{2}}\times 10^{12}~{\rm s}\;.$$

This leads here, for such $B$, to $\tau_{\rm cool}\approx 5$~s at $\nu_{\rm NIR}\approx 10^{14}$~Hz. Sustained injection of particles is therefore required to power the flare during 1--2~h. On top of that, the electron density is so high that synchrotron self-absorption (SSA) comes into play right below $8\times10^{13}$~Hz, with an optically thick power-law $\nu F_{\nu}^{\rm SSA}\propto\nu^{{7}\over{2}}$ \citep{rybicki79}. As a consequence, we cannot observe the power-law with index $(3-p)/2$ for the synchrotron hump in contrast to the inverse Compton one, which also suffers from SSA but only at low frequencies. In any case the expected values for $B$ and $n_{\rm e}$ in the inner accretion flow around the black hole are orders of magnitude smaller ($B\approx10$~G and $n_{\rm e} \approx 10^7$~cm$^{-3}$ for \citet{yuan03}), even though interestingly the magnetic field at equipartition for particles with a typical energy $\gamma_{\rm max}=100$ and density $n_{\rm e}=2.2\times 10^{12}$~cm$^{-3}$ is $B_{\rm eq}\simeq \sqrt{8\pi \gamma_{\rm max} n_{\rm e} m_{\rm e}c^2}\approx7\times10^4$~G, which is almost of the order of magnitude of the SSC magnetic field.

Another weakness of the model is that it predicts in sync variation for the NIR and X-ray light curves, i.e. there should also be visible substructure in the X-ray light curve. 

But an SSC scenario succeeds in explaining the simultaneity of the X-ray and NIR flares as well as the difference in their widths. Indeed, the SSC X-ray flux goes quadradically in $n_{\rm e}$ whereas the NIR flux goes linearly in  $n_{\rm e}$. So, if one suppose the evolution of $n_{\rm e}(t)$ in time has a gaussian profile, then $n_{\rm e}(t)^2$ will have a width $1/\sqrt{2}$ times narrower than $n_{\rm e}(t)$, which could explain the discrepancies of the X-ray and NIR light curves.

We expose here the results for a power-law energy distribution of electrons, but other distributions such as a relativistic Maxwellian (thermal distribution of typical Lorentz factor $\theta_{\rm e}  = \frac{kT_{\rm e}}{m_{\rm e}c^2}$) have been explored in past works \citep[e.g.][]{liu06b}. \citet{dodds09} applied this distribution in an SSC pattern to this flare and found similar results for the physical parameters $B$, $R$, and $n_{\rm e}$. This is because both the power-law and the relativistic maxwellian distributions have a characteristic peak energy, $\gamma_{\rm max}$ or $\theta_{\rm e}$, which determines the relative positions of the synchrotron and inverse Compton bumps.

For the first time, we find it difficult to explain a simultaneous X-ray/NIR flare with SSC emission. As a matter of fact, past observations dealt with weaker flares and poorer spectral information. 
\citet{eckart06a} modeled their contemporaneous X-ray/NIR flares with SSC but had no individual X-ray and NIR indices, so in particular the position of the inverse Compton peak was free, which relaxed the constraints on $B$. In contrast, \citet{marrone08} obtained individual NIR and X-ray spectra, though they were not exactly simultaneous. The flare was fainter than the one presented here and the X-ray spectral index was consequently poorly constrained ($\beta_{\rm X}=0.0^{+1.6}_{-1.0}$). The SED could thus be accommodated with a hard X-ray power-law, which again relaxed the constraints on $B$ and yielded acceptable SSC physical parameters.

\begin{table}
\caption{Parameters of the radiative processes that match the SED of the April 2007 flare from \sgra.}
\centering
\begin{tabular}{llll}
\\
\hline
Parameters&SSC&EC&SB\\
\hline\hline
$p$\dotfill 					& 2 					& 2							& 2--3\\	
$\gamma_{\rm min}$\dotfill 	& 1					& 1							& 1\\
$\gamma_{\rm max}$\dotfill 	& $10^2$				& $10^3$ 			   	 		& $9\times10^4$\\
$\gamma_{\rm br}$\dotfill 		&  --- 				&  --- 						& $9\times10^2$\\
$B$ (G)\dotfill  				& $10^3$  			& 40							& 50\\
$n_{\rm e}$ (cm$^{-3}$)\dotfill	& $2.2\times10^{12}$  	& $>1.8\times10^{10}$			& $7.6\times10^{6}$\\
$R$  (cm)\dotfill			& $1.3\times10^{10}$	& $<1.6\times10^{11}$			& $1.4\times10^{12}$ \\
\hline
\\
\end{tabular}
\label{table}
\end{table}

\subsection{External Compton}

\begin{figure} 
 \centering
 \includegraphics[trim=0cm 8cm 0cm 8.5cm, clip=true, width=14cm]{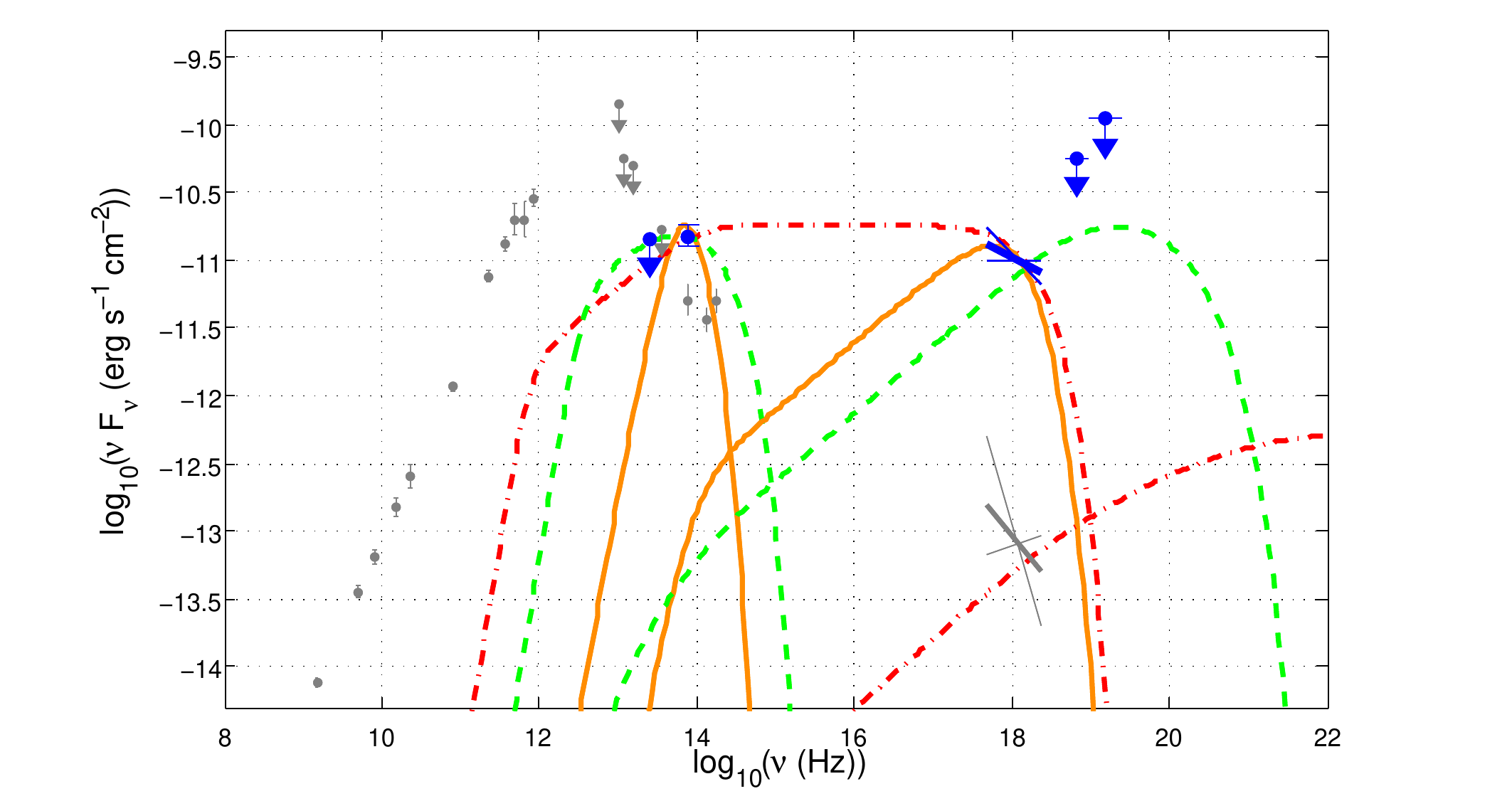}
 \caption{Spectral energy distribution of \sgra. Dark gray measurements correspond to the quiescent state. Radio and submm points are extracted from \citet{zhao01,zylka95,marrone08}, FIR and MIR upperlimits from \citet{telesco96,cotera99,eckart06a,schodel07}, NIR points from \citet{genzel03} and the X-ray bow tie from \citet{baganoff03a}.  The \xmm/EPIC~spectrum of the April 4$^{\rm th}$ flare, the mean \vlt/NACO flux, and the \vlt/VISIR and \integ/ISGRI upper limits are overplotted in blue. The synchrotron and synchrotron self-Compton contributions of the SSC, EC, and SB models are plotted in orange (solid lines), green (dashed lines), and red (dash dotted lines), respectively (see the parameters of the fit in Tab.~\ref{table}).}
\label{sed}
\end{figure}

Another alternative is that transiently accelerated relativistic electrons initiate the NIR flare through synchrotron and upscatter ambient low energy photons to the keV range, thus causing the X-ray flare. The most abundant source of photons around \sgra~are the ones from the submm bump at $\nu_{\rm submm}\approx10^{12}$~Hz (see Fig.~\ref{sed}); we will designate by external Compton (EC) the comptonization of these photons by NIR electrons. We keep the same parametrization as for the SSC case in Sect.~\ref{ssc}. We can estimate the maximal Lorentz factor of the particle and magnetic field as we did in the previous section, by switching respectively $\nu_{\rm max}^{\rm sync}$ and $\nu_{\rm max}^{\rm ic}$, by $\nu_{\rm submm}$ and  $\nu_{\rm X}$ in Eq.~\ref{gamma} and \ref{B}. By this means we find $\gamma_{\rm max}\approx 10^3$ and $B\approx 40$~G. Such a magnetic field is more reasonable than in the SSC picture, the cooling time is $\sim$10~min, and we no longer have to worry about SSA. Besides we know that whenever synchrotron and inverse Compton occur at the same place, the respective luminosities are linked via:

$$\frac{L^{\rm ic}}{L^{\rm sync}}=\frac{U_{\rm seed}}{U_{\rm B}}$$

where $U_{\rm B}=\frac{B^2}{8\pi}$ is the magnetic energy density and $U_{\rm seed}$ is the seed photons energy density. If $A=4\pi R_{\rm Q}^2$ denotes the surface area of the region of particles driving the quiescent submm lumninosity $L^{\rm submm}$, then $U_{\rm seed}=\frac{L^{\rm submm}}{cA}$. Thereby, we can assess the radius of the quiescent emission:

$$R_{\rm Q} \simeq 0.016 \times 10^{12} \left( \frac{  L_{\rm NIR} } { L_{\odot} }  \right)^{1\over2} \left( \frac{  L_{\rm submm} } { L_{\odot} }  \right)^{1\over2} \left( \frac{  L_{\rm X} } { L_{\odot} }  \right)^{-{1\over2}} \left( \frac{  B } { 40~{\rm G} }  \right)^{-1}~{\rm cm}\;,$$

where the solar luminosity is $L_{\odot}=3.8\times10^{33}~{\rm erg\,s}^{-1}$.
The resulting radius is $R_{\rm Q} \approx 0.1~R_{\rm S}$, and this is probably the main weakness of the EC scenario, because VLBI measurements of \sgra~at 1.3~mm give an intrinsic size for the quiescent region of the order of the Schwarzschild radius \citep{doeleman08}. To compute the electron density $n_{\rm e}$ and the size of the flaring region $R$, a detailed treatment of the quiescence spectrum has to be taken into account, as done by \citet{dodds09} with the quiescent RIAF model of \citet{yuan03}. Just to get a feeling of these parameters, we will consider that the flare is embedded in the submm region so that $R<R_{\rm Q}$ and consequently we find that $n_{\rm e}$ must be at least $1.8\times 10^{10}$~cm$^{-3}$ to fit the NIR synchrotron data.

 To complement the study of \citet{dodds09}, we have computed the synchrotron self-Compton emission that will naturally come along in the EC scheme. Interestingly, this contribution peaks in the soft gamma-ray band, less than one order of magnitude below the \textit{INTEGRAL} upper limits (see Fig.~\ref{sed}, dashed line). If this scenario is real, it would tend to flatten the X-ray spectral slope towards high energies, thus making the prospects for future soft gamma-ray missions promising.

As in the SSC scheme, EC ensures simultaneity of the flare in the X-ray and NIR light curves. It could also provide an explanation for the absence of substructures in the X-ray light curve. Indeed, if one naively neglect the synchrotron losses, then the synchrotron luminosity is proportional to $B^{3\over2}$ whereas the inverse Compton luminosity is, {\it a priori}, independent of $B$ because it depends on $L_{\rm submm}$ and not on $L_{\rm NIR}$ as for SSC. So the NIR flare should be subject to $B$ variations contrary to the X-ray flare. A clumpy magnetic field in which the flaring region moves is maybe the key to these observed or unobserved features.   

\citet{yusef-zadeh06a} proposed that their synchronous observation of a flare with the \hst~and \xmm~in 2004 resulted from EC with an acceptable size of the submm quiescence region of $\sim$$10~R_{\rm S}$. This was possible only because the spectrum of the X-ray flare was hard, $\beta_{\rm X}\approx0.5$ \citep{belanger05}, and therefore allowed for a larger $\gamma_{\rm max}$, a lower $B$ and larger $R_{\rm Q}$. These authors also pointed out that the quiescent electrons reponsible for the submm bump were also likely to upscatter the NIR flare photons to keV energies as well, even though this may be a second order effect \citep{dodds09}.

\subsection{Synchrotron with break}
From our flare observations, it is clear that $\beta_{\rm NIR} \neq \beta_{\rm X}$ so that we cannot fit the entire SED of the flare with a single synchrotron power-law. But broad synchrotron power-laws are known to exhibit breaks of several kinds. In particular, a natural break comes from the synchrotron cooling of the electrons, which generates a difference of slopes of $|\Delta p |=1$ in the electrons distribution (and $|\Delta \beta |=0.5$ in the photons distribution) between the power-law below and above the break. This simple ``synchrotron with a break'' model (SB) would actually suit our measurements. The frequency at which the break is supposed to occur depends upon the modeling adopted. Here, we assume a ``leaking box'' model in which a constant injection of fresh particles is balanced by the synchrotron cooling of this electrons on a timescale $\tau_{\rm cool}$ and their escape on another typical timescale. For the latter variable, we will take the the dynamical timescale $\tau_{\rm dyn}=\sqrt{\frac{r^3}{2GM_{\bullet}}}\approx 5$~min, where $r=3~R_{\rm S}$ is the radius of the last stable orbit for a non spinning black hole. The condition $\tau_{\rm cool} = \tau_{\rm dyn}$ provides the frequency of the spectral break:

\begin{equation}
\label{break}
\nu_{\rm br}=6.37 \left(\frac{B}{1~\rm{G}}\right)^{-3}\times10^{18}~{\rm Hz}\;.
\end{equation}

In Tab.~\ref{table} we list a sketch of physical parameters for SB that match the SED well. Again we chose $\gamma_{\rm min}=1$, and $\gamma_{\rm max}=9\times10^4$ to engender photons up to the X-ray range. $B$ was chosen to satisfy Eq.~\ref{break} for a break at $\sim$$10^{14}$~Hz. Finally, $n_{\rm e}$ and $R$ were adjusted to normalize the spectrum with reasonable values and in order not to violate our $\gamma$-ray constraints with the natural SSC component (see the rising SSC hump in red on bottom right of Fig.~\ref{sed}) coming along with synchrotron radiation. Note once more on Fig.~\ref{sed} the SSA below $\sim$$10^{12}$~Hz.

SB is appealing because, compared to SSC and EC, it yields less extreme values in terms of $B$, $n_{\rm e}$, and $R$. However, in the SB case, there is no obvious justification for the differents durations of the flare in X-rays and NIR, and the presence/absence of substructures in the light curves. Here, we only discuss the average spectrum of the flare, where SB certainly requires a more detailed examination of the time evolution of the phenomenon, beyond the scope of this paper.

\section{Conclusions}

This paper complements a series of articles about the April 2007 synchronous observations of the Galactic Center from radio to $\gamma$-rays \citep{porquet08,dodds09,yusef-zadeh09}.
Here, we have recapped the results on the brightest flare ever detected simultaneously at NIR and X-ray frequencies. 
We have also reported for the first time $\gamma$-ray constraints on such an event, which, added to our MIR/NIR/X-ray spectral measurements, constitute the broadest simulaneous spectrum of a flare ever achieved.
The essential observational conclusions may be summarized as follows:
\begin{itemize}
\item the peaks of the X-ray and NIR emissions are coincident within 3~min;
\item the width of the NIR flare light curve is broader than the X-ray one by a factor $\sim$2;
\item the NIR light curve is substructured on a timescale of $\sim$20~min while the X-ray light curve is rather smooth;
\item there is no detectable MIR counterpart;
\item the soft $\gamma$-ray source \igr~is non variable.
\end{itemize} 
The high quality of the spectral information we gathered allowed for a discussion of the several classical radiative processes models employed to explain the flares: SSC, EC, and SB. Yet, none of these mechanisms is entirely satisfactory to meet our observations. The theoretical inquiries to come will have to take into account the time evolution of the phenomenon and the aging of the radiating particles to better connect the light curves and spectra. From an observational stand point, it will be useful to repeat such NIR/X-ray measurements in a near future to get two respective individual and fully contemporaneous spectra, which has never been accomplished thus far. As we have seen, one key probe of what powers the flares, is a better determination of the X-ray spectral slope. In a more distant future, thanks to a broad X-ray sensitivity over the 1--80~keV band and a high angular resolution above 10~keV, \simx~should address this issue and resolve the GC region in soft $\gamma$-rays.


\section*{Acknowledgements}

GT acknowledges  M. Falanga, D. G\"otz, and J. Chenevez for help with the \integ~data analysis, Y. Cl\'enet, B. Draine, and M. Morris for useful discussions about NIR/MIR measurements, and CEA Saclay for financial support to attend the 37$^{\rm th}$ COSPAR meeting in Montreal. 
 
Part of this work has been funded by the french Agence Nationale pour la Recherche through grant ANR-06-JCJC-0047. At Arizona, this work was also supported by NASA grants NNX08AX33G and NNX08AX34G.
      
\integ~is an ESA project with instruments and science data center funded by ESA member states (especially the PI countries: Denmark, France, Germany, Italy, Switzerland, and Spain), the Czech Republic, and Poland, and with the participation of Russia and the US. 

The \xmm~project is an ESA Science Mission with instruments and contributions directly funded by ESA Member States and the USA (NASA).


\end{document}